\documentclass[aps,floatfix,epsfig]{revtex4}
\usepackage{pdfpages}
\usepackage{subfigure}
\usepackage{amsmath}
\usepackage{eqnarray,amsmath}
\usepackage{epstopdf}
\usepackage{mathrsfs}
\usepackage{natbib}
\usepackage{amssymb,latexsym}
\usepackage{dcolumn}
\usepackage{graphicx}
\def\be{\begin{equation}}
	\def\ee{\end{equation}}
\def\ba{\begin{eqnarray}}
	\def\ea{\end{eqnarray}}

\begin{document}
\title{Point mass Cosmological Black Holes}

\author{Javad T. Firouzjaee}
\affiliation{ School of Astronomy, Institute for Research in Fundamental Sciences (IPM), P. O. Box 19395-5531, Tehran, Iran}
\email{j.taghizadeh.f@ipm.ir, tohid.fagihi@gmail.com}
\author{Touhid Feghhi}
\affiliation{ School of Astronomy, Institute for Research in Fundamental Sciences (IPM), P. O. Box 19395-5531, Tehran, Iran}
%
\begin{abstract}

\textbf{Abstract:} Real black holes in the universe are located in the expanding accelerating background which are called the cosmological black holes. Hence, it is necessary to model these black holes in the cosmological background where the dark energy is the dominant energy. In this paper, we argue that most of the dynamical cosmological black holes can be modeled by point mass cosmological black holes. Considering the de Sitter background for the accelerating universe, we present the point mass cosmological background in the cosmological de Sitter space time. Our work also includes the point mass black holes which have charge and angular momentum. We study the mass, horizons, redshift structure and geodesics properties for these black holes.

\end{abstract}
%
%

\maketitle

\tableofcontents
\section{Introduction}

After Hubble discovery of the cosmic expansion, people needed to make black hole models which are embedded in the expanding Friedmann-Lema\^{i}tre-Robertson Walker background. The 1933 McVittie solution \cite{McVittie:1933zz} was the first attempt to model the point mass black hole in the cosmological background.  The question arises to whether the effects of the cosmological expansion on gravitating local systems, such as the solar or galaxy system, accompanying
the expansion of the universe. The McVittie black hole is
embedded in a general FLRW background, so that the region between the black hole horizon and the cosmological horizon is evolving. Even though this model was comprehensively studied in different aspects of black hole \cite{Nolan:1999kk,Kaloper:2010ec,Lake:2011ni},  the McVittie metric cannot describe a black hole
evolving in an FLRW universe. The collection
of known exact solutions with respect to the point-mass modeling were investigated in \cite{Point mass-CBH}.\\

The main criticism for some these metrics is that they manufacture the geometry and attribute it to the black hole models. For example, if one replaces the constant mass in the Schwarzschild metric, $m$, with a specific function of $m(t,r)$, the energy condition might be violated. With this action, matter on the right-hand side of the Einstein equation will be restricted, and the black hole horizon and singularity change their position and properties generally. The standard way to derive a metric is to know its matter field and its symmetry. Along this way Lema\^{i}tre-Tolman-Bondi (LTB) model was introduced which describe the perfect fluid collapse in the spherically symmetric space time \cite{LTB}. Apart from its dynamical nature, the FLRW is a special case of this metric and can be modeled as a background. Since the geometry is not static, the need for a local definition
of black holes and their boundaries (horizons) have led to concepts such as an isolated horizon \cite{Ashtekar:2000sz}, Ashtekar and Krishnan’s dynamical horizon \cite{Ashtekar:2002ag}, and Booth and Fairhurst’s slowly evolving horizon \cite{Booth:2003ji}. Inspired by these LTB metric properties, the cosmological black hole (CBH) can be built  \cite{Firouzjaee-2010,Firouzjaee-penn,Moradi-2015} that its singularity and horizon is formed during the collapse. \\

In many respects, a CBH shows different properties relative to the stationary black holes. The first evident difference is that in the CBH has one more horizon except black hole horizons (event and apparent horizon) which called the cosmological horizon \cite{Gibbons:1977mu,book-faraoni,Helou:2016xyu}, and the causal structure of the black hole will be different from stationary one. Second, the mass definition of the black hole will need to be extended to the quasi-local masses \cite{Firouzjaee:2010ia} rather than like a ADM mass. Even though the Hawking radiation from the stationary black hole is revised for CBH case \cite{radiation-CBH}.

The first CBH application is in the primordial black hole modeling \cite{PBH-CBH}. These black holes form when the FLRW background density perturbations exceed some threshold values in the radiation dominated era. One can use McVittie CBH  to model the cosmological defect in the inflationary phase \cite{McVittie-inflation}. The second CBH application is to model the structure in the matter dominated era \cite{CBH-structure formation} where these models were used to investigate the dark energy, virialization, rotation curve and week-lensing, etc.  \\

In this paper, in section II we infer the point mass CBH form a big class of CBH which can be used for the cosmological structure. Section III, consider the validity of the point mass approximation for astrophysical black holes.  IV, V and VI is devoted to make  a point mass CBH in presence of mass, mass-charge and mass-angular momentum. Then, in Section VI we study these black holes mass, horizon, redshift structures and geodesics properties. The conclusion and discussions are given in Section VIII.

\section{From dynamical black holes to the point mass black holes}

The main feature of a dynamical black hole in astrophysics is its matter flux which helps the black hole formation and its growth. It was shown \cite{Firouzjaee:2010ia,Moradi-2015} that the black hole apparent horizon growth is proportional to the matter flux which falls to it and eventually tends to the black hole event horizon when all matter flux is absorbed by black hole.
Here we are interested in the case that the black hole evolves slowly due to the matter flux. First we introduce the dynamical horizon properties \cite{Ashtekar:2002ag} which is a general case slowly evolving horizon \cite{Booth:2003ji} and then we consider the slowly evolving limitation. Geometry of the dynamical horizon $H$ is expressed by the unit
normal to $H$ by $\hat{\tau^a}$; $g_{ab}\hat{\tau^a}\hat{\tau^b}=-1$. The unit space-like vector orthogonal to $S$ (a point on apparent horizon which has 2-sphere topology) and tangent to $H$ is
represented by $\hat{r^a}$. The rescaling freedom
in the choice of null normals will be fixed via $l^a=\hat{\tau^a}+\hat{r^a}$ and
$n^a=\hat{\tau^a}-\hat{r^a}$. We introduce the area radius $R$, a
function  which is constant on each $S$ and satisfies $a_S = 4\pi
R^2$.  Now, the 3-volume $d^3V$ on $H$ can
be decomposed as $d^3V = |\partial R|^{-1}dR d^2V$ where $\partial$
denotes the gradient on $H$. Hence, as we will see, our
calculations will simplify if we choose $N_R = |\partial R|$ which is amplitude of normal vector of the $R=const$ surface.

We define the flux of energy associated with $\xi_{(R)}^a = N_R \ell^a$ across $\Delta H$ as:
\begin{eqnarray} \mathcal{F}^{(R)}_{\rm matter} := \int_{\Delta H}
	T_{ab}\hat{\tau^a}\xi_{(R)}^b d^3V=\nonumber\\
	\frac{1}{G}(M(r_2)-M(r_1)).
\end{eqnarray}
In the spherically symmetric case $M(r)$ is the Misner-Sharp mass \cite{Firouzjaee:2010ia}.
With calculating this quantity on the apparent horizon
\be
\frac{dM}{dt}|_{AH}=(M'\frac{dr}{dt}+\dot{M}).
\ee

Here $\dot{}$ and $'$   are partial differentials relative to $t$ and $r$ respectively. Take a collapsing ideal fluid within a compact spherically symmetric
spacetime region described by the following metric in the comoving
coordinates $(t,r,\theta,\varphi)$:
\begin{equation}
ds^{2}=-e^{2\nu(t,r)}dt^{2}+e^{2\psi(t,r)}dr^{2}+R(t,r)^{2}d\Omega^{2}.
\end{equation}
Assuming the stress-energy tensor for the perfect fluid in the
form
\begin{eqnarray}
T^{t}_{t}=-\rho(t,r),~~T^{r}_{r}=p_{r}(t,r),~~\nonumber\\ T^{\theta}_{\theta}=
T^{\varphi}_{\varphi}=p_{\theta}(t,r)=w \rho(t,r),
\end{eqnarray}
with the weak energy condition
\begin{equation}
\rho\geq0~~\rho+p_{r}\geq0~~\rho+p_{\theta}\geq0,
\end{equation}
where $w$ describes the equation of state which is a barotropic equation of state between energy density, $\rho$, and pressure, $p$. 
Einstein equations give,

\ba \label{gltbe2}
\rho=\frac{2M'}{R^{2}R'}~,~~p_{r}=-\frac{2\dot{M}}{R^{2}\dot{R}},
\ea
where $M$, Misner-Sharp mass, is defined by

\ba \label{gltbe3} e^{-2\psi}(R')^{2}-e^{-2\nu}(\dot{R})^{2}=1-\frac{2M}{R}. \ea
If a collapsing metric is build by this metric, one can show that the apparent horizon will form at $R=2M$ surface \cite{Firouzjaee-penn}. We can define the matter flux into the apparent horizon matter flux relative to fluid 4-velocity 
\be
\label{flux}
D_t M= \frac{1}{e^\nu}\frac{dM}{dt}|_{AH}=\frac{1}{e^\nu}(M'\frac{dr}{dt}+\dot{M}).
\ee

Using the Einstein equation \cite{Firouzjaee-penn,Helou:2016xyu}  we get
\be
D_t M=4\pi R_{H}^2 (-U)\frac{\rho+p}{1-8\pi R_{H}^2 \rho}
\ee
where $U=\dfrac{\dot{R}}{e^\nu}$ and the $R_H$ the areal radius on the apparent horizon. At first glance,  we can see that the matter (flux) on the apparent becomes zero when density becomes zero on the apparent horizon. We define this quantity which characterize the slowly evolving horizon \cite{Booth:2003ji} when
\be
\epsilon^2=\alpha  \theta_n^2R_{H}^2 \ll 1,
\ee
where $ \theta_n $ is the ingoing null geodesics expansion and alpha is a constant. This quantity in the spherically symmetric space time becomes \cite{Firouzjaee-penn}
\be \label{epsilon}
\epsilon^2= 4\pi R_{H}^2 \frac{(1+w)\rho}{1-4\pi R_{H}^2 (1-w)\rho}
\ee
To have an slowly evolving black hole with mass $M=2R_H$ the energy density for this black hole, $\rho$, must be small. In the special case the $\epsilon = \rho=0$ we get the isolated horizon \cite{Ashtekar:2000sz} which the matter flux is zero. \\

\textit{Proposition:} If a dynamical black hole evolves and finally its horizon becomes isolated, the space time geometry around the black hole horizon becomes Schwarzschild metric with the same Misner-Sharp mass.\\

If the horizon becomes isolated i.e. $\epsilon =0$ then from Eq. (\ref{epsilon}) the energy density becomes zero, $\rho=p=0$. Hence we have underdensity (vacuum) around the black hole horizon. On the other hand from uniqueness theorem we know that the vacuum solutions of the Einstein equation around a spherically symmetric mass distribution have Schwarzschild form
\be
ds^2=-(1-\frac{2C}{R})dt^2 + (1-\frac{2C}{R})^{-1}dR^2+R^2 d\Omega^2.
\ee
Using the fact the $M=2R_H$ is the apparent horizon for Schwartzchild metric (as boundary condition), we get $C=M$. Therefore, in the isolated horizon case the space time geometry around horizon becomes Schwarzschild with the  Misner-Sharp mass, $C=M=\frac{R_H}{2}$.\\ 

 Located in the cosmological background, the CBH collapsing part separates from the expanding part and the density between them decreases. This underdense region is usually called void in the cosmology. As a result, the matter flux can not exist  forever and after sometime it decreases \cite{Firouzjaee:2010ia,Moradi-2015}. In this case as inferred above taking the black hole as a point mass is a good model.\\

\section{What is the dynamical phase of the   Sgr A$^*$ and M87}
To know how the point mass approximation is correct for the real astrophysical black hole, let us study the dynamical growth of the two famous black holes Sgr A$^*$ and M87.
Sgr A$^*$ is a supermassive black hole in our galaxy's center. From the radio source of the galactic center,  the Sgr A$^*$  mass is estimated $ M \sim  10^{6} M_{\odot} $. From the the X-ray and infrared emission of the
Galactic center \cite{sgr}, one  provides an upper limit on the Sgr A$^*$ black hole mass accretion rate as
$ \dot{M}< 10^{-5} M_{\odot} yr^{-1} $. For an astrophysical event such as lensing for light the typical lenght (time) scale is about $kpc\sim 10^{19} m$ which is equivalent to $10^3 yr$ for passing light of this length \cite{Ghez:2008ms}.  The black hole mass growth for this time scale is $\delta M < 10^{-2} M_{\odot}$. Since the black hole mass is proportional to its radius the change of the black hole horizon becomes
\be
\frac{\delta R_H}{R_H}= \frac{\delta M }{M}  < 10^{-8}.
\ee
As a result, in practice the black hole remains static relative to the astrophysical time scale and we can make the point mass approximation in many cases for the black holes physics study as the astrophysical text book do \cite{bh-book}.\\

Another supermassive black hole is M87 in cluster Virgo fifty million light-years away which is the most massive black holes known and has been the subject of several stellar and gas-dynamical mass measurements. This suppermassive black hole mass is 
 $ M \sim 10^{9} M_{\odot}$ and the mass accretion rate of this black hole is approximated by $ \dot{M}< 10^{-4} M_{\odot} yr^{-1} $ \cite{m87}. The typical distance of the Virgo cluster from us is about $10^{22} m$  which is equivalent to $10^6 yr$ for coming light. At this time scale the black hole mass growth will be $\delta M < 10^{2} M_{\odot}$ and equivalently the the black hole horizon growth becomes
 \be
 \frac{\delta R_H}{R_H}= \frac{\delta M }{M}  < 10^{-7}.
 \ee
 Similar to the Sgr A$^*$, in practice this black hole remains static relative to the astrophysical time scale  \cite{Owen:2000vi} and we can make the point mass approximation in many cases for the black hole's physics study.\\

\section{Point mass CBH }

In the last section we inferred that since the matter and matter flux around the black hole decrease and the black hole mass is bigger than the total matter around it, the point mass black hole can be a good model for CBH. On the other side, the cosmological constant is the best model to describe the cosmic acceleration which we called it dark energy in the matter sector. The de Sitter metric is the Einstein equation solution with the cosmological constant. Our analysis based on two paradigm: first, the dark energy is the main cosmological matter (about 70 \%) and all other matter (dark matter and baryonic  matter) located in the matter flux which have fallen in the black hole; second at last time of evolution the matter flux becomes zero \cite{Firouzjaee-2010,Moradi-2015,Firouzjaee-penn} and we get a point mass black hole. In this case, it is sufficient to find the point mass black hole in the cosmological de Sitter background. \\

One way to make a cosmological black hole is to add the FLRW scale factor as a conformal coefficient for the point mass black hole metric. Sultana and Dyer \cite{sultana} have introduced a metric which FLRW scale factor is a conformal coefficient of the Schwarzschild metric. However that metric violate the energy condition and does not describe the ordinary matter. To do similar analysis with the de Sitter line element \cite{Culetu} consider the de Sitter metric 
 
\begin{equation}
\label{bb1}
ds^{2} = -d \tau ^{2} + e^{2 \sqrt{\frac{\Lambda}{3}} \tau} (dr^{2} + r^{2}d \Omega^{2} ).
\end{equation}
By the coordinate transformation $ t = \sqrt{\frac{\Lambda}{3}} e^{-\sqrt{\frac{\Lambda}{3}} \tau} $, the metric (\ref{bb1}) will be transformed into conformally flat form 
\begin{equation}
ds^{2} = \frac{\Lambda}{3t^{2}}[-dt^{2} + dr^{2} +r^{2} d \Omega ^{2} ].
\end{equation}
If we embed a black hole in this space time and write the conformal Schwarzschild metric, we will have
\begin{equation}
ds^{2} =  \frac{\Lambda}{3t^{2}}[-(1-\frac{2m}{r})dt^{2} +\frac{1}{(1-\frac{2m}{r})} dr^{2} + r^{2}d \Omega ^{2} ].
\end{equation}  
By calculating the Einstein tensor for this line element we see that the strong energy condition is violated, so this metric is not proper for our study and instead we consider the general form of Schwarzschild de Sitter metric.\\ 

One may think that the Schwarzschild-de Sitter is a good metric for the CBH, but Schwarzschild-de Sitter has been written in the static coordinate. Consequently it is needed find the Schwarzschild-de Sitter metric in the cosmological coordinate to present the point mass CBH.
Since the standard cosmological metrics are written in the synchronous coordinate, we first have to transform the Schwarzschild-de Sitter to the synchronous coordinates.
We know that the  Schwarzschild-de Sitter metric is given by 
\begin{equation}
\label{a1}
ds^{2} =  - \Phi   dt^{2}  + \Phi ^{-1}  dR^{2} + R^{2} d \Omega ^{2},
\end{equation}

where

\begin{equation}
\label{s-d}
\Phi =   1- \frac{\Lambda}{3} R^{2} - \frac{2M}{R}.
\end{equation}

By these  coordinate transformations  
  \cite{podolski}

\begin{equation}
\label{a2}
\begin{split}
  &  d \tau = dt - \frac{\sqrt{1 - \Phi}}{\Phi}dR,
\\& dr = -dt +  \frac{1}{\Phi \sqrt{1- \Phi}} dR,
\end{split}
\end{equation}
 metric  (\ref{a1})  will be 

\begin{equation}
\label{ }
ds^{2} = -d \tau ^{2} + (\frac{2M}{R} +  \frac{\Lambda}{3}R^{2} )dr^{2} + R^{2}d\Omega^{2}.
\end{equation}
 To find $R$ as a function of $r$ and $ \tau $ we can use
\begin{equation}
\label{ }
\int d \tau + d r   = \int \frac{dR}{\sqrt{1-\Phi}}  . 
\end{equation}
Therefore, we can write
\begin{equation}
\label{ }
\tau + r = \int \frac{dR}{\sqrt{\frac{\Lambda}{3}R^{2} + \frac{2M}{R}}} =  \frac{2}{\sqrt{3 \Lambda}} \ln( \Lambda R^{\frac{3}{2}} + \sqrt{6M \Lambda + \Lambda^{2}R^{3}}).
\end{equation}
 
Hence, we can write  $R$

\begin{equation}
\label{ }
 R=   \frac{e^{ - \sqrt{\frac{\Lambda}{3} } (r + \tau) } (e^{\sqrt{3 \Lambda} (r+ \tau ) } -6\Lambda M)^{\frac{2}{3}}   }{2^{\frac{2}{3}} \Lambda^{\frac{2}{3}}}.
 \end{equation}
In the limit where the black hole mass tend to  zero, the metric
  $\displaystyle{\lim_{M \to 0} ds^{2} }$ will be

\begin{equation}
\label{a20}
\begin{split}
ds^{2}   
= -d \tau^{2}  +  \frac{e^{2 \sqrt{\frac{\Lambda}{3}} \tau} }{(2 \Lambda)^{\frac{4}{3}}} [\frac{\Lambda}{3} e^{2 \sqrt{\frac{\Lambda}{3}  }r } d r^{2} +  e^{2 \sqrt{\frac{\Lambda}{3}  }r } d \Omega ^{2} ].
\end{split}
\end{equation}
By redefining $  S =  \frac{e^{2 \sqrt{\frac{\Lambda}{3}  }r } }{(2\Lambda)^{2/3}} $ we get
 
    \begin{equation}
\label{ }
ds^{2}   = -d \tau^{2}  +  e^{2 \sqrt{\frac{\Lambda}{3}} \tau}  [ d S^{2} +  S^{2} d \Omega ^{2} ],
\end{equation}
which is de Sitter metric.\\
With calculating the Ricci scalar, we can see that  the singularity is located at $  r + \tau   =  \frac{1}{\sqrt{3 \Lambda}} \ln(6 \Lambda M) $ that is equivalently $r= 0$.\\

  The Penrose-Carter diagram of these black holes  can be seen in the \cite{Gibbons:1977mu}. 
These black holes properties will be discussed in Section VI.\\

\section{Charged Black hole}

It is usually assumed that the charged black hole can not appear during the gravitational collapse. But a highly magnetized plasma accretes onto the black hole, the charge to mass
ratio can be big in some case. In particular, in the merging of a binary system of neutron stars, it is expected at
the final steps of a gravitational collapse to a charged black hole \cite{ccbh}. 
If we add charge to a point mass we get the  Reissner Nordstrom solution.  If we solve the Einstein equation with cosmological constant and a point mass with charge we get the de Sitter-Reissner Nordstrom solution. Here we use  Carter's  spherically symmetric    three parameter (M,$ \Lambda $,Q)  solution to Einstein's equations  where  Q  is the electric charge of black hole  \cite{carter}. The metric in static coordinates is 
    
    \begin{equation}
\label{a3}
ds^{2}  = -\Phi dt^{2} + \Phi^{-1} dR^{2} + R^{2} d \Omega^{2} ,
\end{equation}

where 

\begin{equation}
\label{ }
\Phi =   1- \frac{\Lambda}{3} R^{2} - \frac{2M}{R} +\frac{Q^{2}}{R^{2}}.
\end{equation}

By coordinate transformations given by (\ref{a2}), the metric (\ref{a3}) will be 

\begin{equation}
\label{a31}
ds^{2}  = - d \tau ^{2} + ( \frac{\Lambda}{3} R^{2} + \frac{2M}{R} - \frac{Q^{2}}{R^{2}} ) dr^{2} + R^{2}d \Omega^{2}.
\end{equation}
We can write
\begin{equation}
\label{a40}
\tau + r = \int  \frac{dR}{\sqrt{\frac{\Lambda}{3}R^{2} + \frac{2M}{R} -\frac{Q^{2}}{R^{2}} }}  .
\end{equation}
If we define $ G(R) =   \int  \frac{dR}{\sqrt{\frac{\Lambda}{3}R^{2} + \frac{2M}{R} -\frac{Q^{2}}{R^{2}} }}      $
 then $  R = G^{-1}(\tau + r)  $. We can calculate the inverse function numerically, as if we draw the function$r+ \tau = G(R)$ and change the variables $(R,G(R))$ to $(G(R),R)$ and then use numerical methods to find the equation of drawn line.
 
Because of existence of three free parameters      $ \Lambda $ ,    $ Q $  and $ M $,  this integral requires tedious calculations (if an analytical solution exists), and even after finding the integral, it would be much harder to find the inverse function $ R $, which will give the solution. Hence, we need to  suppose some simplifying assumptions. At large '$ R $' where 
$ \frac{\Lambda}{3}R^{2}  \gg  \frac{2M}{R}  $ and $ \frac{\Lambda}{3}R^{2}  \gg    \frac{Q^{2}}{R^{2}} $  or equivalently   $  R \gg  Max( \sqrt[3]{\frac{6M}{\Lambda}} ,  \sqrt[4]{\frac{3Q^{2}}{\Lambda}}) $
   
 \begin{equation}
 \begin{split}
\label{}
 \tau + r       & =  \int \frac{dR}{\sqrt{\frac{\Lambda}{3}*R^{2}  +  \frac{2M}{R} -\frac{Q^{2}}{R^{2}} }} 
  \simeq    \sqrt{ \frac {3}{\Lambda} }    \int R^{-1}[1- \frac{3M}{\Lambda R^{3}}  + \frac{3Q^{2}}{2 \Lambda R^{4}}  ] dR   \\ & =   \sqrt{ \frac {3}{\Lambda} }  (\ln (R)  +  \frac{M}{\Lambda R^{3} }  - \frac{3 Q^{2} }{8 \Lambda R^{4}} ).
 \end{split}
 \end{equation}
As a result 
 
 \begin{equation}
\label{a30}
   R =  \exp^{ x_{i}}.
\end{equation}  
Where $ x_{i} $ represents the roots of equation
\begin{equation}
\label{ }
 8(r+ \tau) -  \sqrt{\frac{3}{\Lambda}}(\frac{8 \Lambda x e^{4x} + 8M e^{x} - 3Q^{2} }{\Lambda e^{4x}}) =0
\end{equation}
  For $\displaystyle{\lim_{Q,M \to 0} ds^{2} }$, according to (\ref{a30}), $ R= \exp^{\sqrt{\frac{\Lambda}{3}}(r + \tau ) } $ and  the line element (\ref{a31}) will be  (\ref{a20}) and it represents  de Sitter metric.\\

It might be interesting to  consider the case  $M=0$ (Q $\neq 0$)
\begin{equation}
\label{ }
\int \frac{dR}{\sqrt{\frac{\Lambda}{3} R^{2} -  \frac{Q^{2}}{R^{2}}}}  =  \frac{1}{2 \sqrt{\frac{ \Lambda}{3}}} \ln (\Lambda R^{2} + \sqrt{\Lambda^{2}R^{4} - 3 \Lambda Q^{2}} ) .
\end{equation}

Then

\begin{equation}
\label{ }
R =\frac{e^{-\sqrt{\frac{\Lambda}{3}}(r + \tau)}  \sqrt{e^{4 \sqrt{\frac{\Lambda}{3}} (r + \tau)} +  3  \Lambda Q^{2}}  }{\sqrt{2 \Lambda}} 
\end{equation}

The singularities are located at $ R=0 $ and $ R = \sqrt[4]{\frac{3Q^{2}}{\Lambda} } $.

For the region $  \frac{Q^{2}}{2M} \ll R \ll \sqrt[3]{\frac{6M}{\Lambda}} $  ( if $ \sqrt[3]{\Lambda} Q^{2} \ll M $ ), the equation(\ref{a40}) will be

\begin{equation}
\label{ }
\begin{split}
r+\tau =  &  \int  \frac{dR}{\sqrt{\frac{\Lambda}{3}R^{2}  +  \frac{2M}{R}  - \frac{Q^{2}}{R^{2}}}}  
= \int \frac{\sqrt{R}}{\sqrt{2M}} [1 - \frac{Q^{2}}{2MR}  +  \frac{\Lambda}{6M}R^{3} ]^{- \frac{1}{2}}dR \\ & \simeq   \frac{1}{\sqrt{2M}} ( \frac{2}{3}R^{\frac{3}{2}}  + \frac{Q^{2}}{2M}R^{\frac{1}{2}}  - \frac{\Lambda}{54M}R^{\frac{9}{2}} ) 
\end{split}
\end{equation}

 and we will have $ R = x_{i} $, where $ x_{i} $ are the roots of this equation
 \begin{equation}
\label{ }
-5832M(r+ \tau)^{2} +  729 M^{2}Q^{2} x + 1944MQ^{2}x^{2} + 1296 x^{3} - 54\Lambda M^{2}Q^{2} x^{5}  - 72 \Lambda M x^{6} + \Lambda ^{2} M^{2} x^{9} = 0
\end{equation}

  The Penrose-Carter diagram of these black holes can be seen in the \cite{Belhaj:2009ii}. 
  \\
\section{Kerr- de Sitter Black hole}

In this section  we want to describe the  point mass CBH with the angular momentum. In Boyer-Lindquist  like coordinates employed by Carter, the Kerr-de Sitter line element  will be  \cite{carter}:

\begin{equation}
\label{a7}
\begin{split}
ds^{2} =  &  (R^{2} +a^{2}cos^{2} \Theta  )[ \frac{dR^{2}}{\Delta_{R}}  +  \frac{d \Theta ^{2} }{1 + \frac{\Lambda}{3} a^{2} cos^{2} \Theta } ]   +  sin^{2} \Theta \frac{ (1 +\frac{\Lambda}{3}a^{2}cos^{2} \Theta  ) }{R^{2} + a^{2} cos^{2} \Theta}[\frac{adt - (R^{2}+a^{2})d \varphi }{1 + \frac{\Lambda}{3} a^{2}}]^{2}   \\ &   - \frac{\Delta_{R} }{(R^{2} +a^{2}cos^{2} \Theta  )} [\frac{dt - a sin^{2} \Theta d \varphi  }{1 + \frac{\Lambda}{3} a^{2}}]^{2},
\end{split}
\end{equation}

where 'a'  is Kerr parameter and 

\begin{equation}
\label{ }
  \Delta_{R} =  (R^{2} + a^{2} ) (1  - \frac{\Lambda}{3} R^{2}) - 2MR.
\end{equation}

The Kerr-de sitter metric is rather complicated , so finding a coordinate in which metric becomes in the form of $ ds^{2} = \Phi dt^{2}  - \Phi ^{-1} dR^{2}   - R^{2} d \Omega ^{2}$  is elusive. In this way,  to simplify metric, we first take $ \Theta = 0 $ polar cut of the metric and due to axisymmetry  $  \varphi $ can be set to equal any value from 0 to $2 \pi$. With these assumptions the metric (\ref{a7}) will be

\begin{equation}
\label{ }
ds^{2}  = \frac{R^{2} + a^{2} }{\Delta_{R} }dR^{2}  - \frac{\Delta_{R} }{R^{2} +a^{2}} [\frac{dt}{1 +  \frac{\Lambda}{3}a^{2}} ]^{2}.
\end{equation}
 If we suppose $dt^{'} = \frac{dt}{1 +  \frac{\Lambda}{3}a^{2}}  $  we will have

  \begin{equation}
\label{ax}
ds^{2}  = \frac{R^{2} + a^{2} }{\Delta_{R} }dR^{2}  - \frac{\Delta_{R} }{R^{2} +a^{2}} (dt^{'})^{2}.
\end{equation}

Similar to the last section  if we use coordinate transformations (\ref{a2})  the line element (\ref{ax})  will be
 \begin{equation}
\label{ }
ds^{2}  = -d \tau^{2} + ( \frac{\Lambda}{3}R^{2}  + \frac{2MR}{R^{2}+a^{2}} )dr^{2} ,
\end{equation}
 Where 
\begin{equation}
\label{ }
\int d \tau + d r   = \int \frac{dR}{\sqrt{1-\Phi}}  , 
\end{equation}
And

\begin{equation}
\label{ }
\Phi =   \frac{\Delta_{R} }{R^{2} +a^{2}} .
\end{equation} 

 We can write
\begin{equation}
\label{ }
\tau + r = \int   \frac{dR}{\sqrt{1- \frac{\Delta_{R} }{R^{2} +a^{2}}  }} =   \int \frac{dR} {\sqrt{\frac{\Lambda}{3}R^{2}  + \frac{2MR}{R^{2}+a^{2}} }} .
\end{equation}

Since this integral calculation is complex, we will consider two simple cases $ R \ll a    $  and $ R \gg a  $. \\

 For $ R \ll a $ we will have
 \begin{equation}
\label{ }
\tau + r    
 \simeq     \int \frac{dR}{\sqrt{\frac{   \Lambda}{3}R^{2}  + \frac{2MR}{a^{2}}    }  }  
\end{equation}

Hence we get

 \begin{equation}
\label{ }
\tau + r =    \sqrt{\frac{3}{\Lambda}}  ln  (\sqrt{\frac{\Lambda}{3}}R +  \frac{M}{a^{2}}  \sqrt{\frac{3}{\Lambda}}     +  \sqrt{-( \frac{M}{a^{2}}   \sqrt{\frac{3}{\Lambda}} )  ^{2}  +( \sqrt{\frac{\Lambda}{3}}R +  \frac{M}{a^{2}}  \sqrt{\frac{3}{\Lambda}}       )^{2}    }).
\end{equation}

To find '$ R $' we must   find the inverse form  of  above function, so it will be

\begin{equation}
\label{ }
R=   \frac{\sqrt{3}}{2 \Lambda}  ( exp( \sqrt{\frac{\Lambda}{3}} (r+ \tau) )  - 2 \frac{M}{a^{2}}  \sqrt{\frac{3}{\Lambda}}    +   ( \frac{M}{a^{2}}   \sqrt{\frac{3}{\Lambda}} )  ^{2}  exp(- \sqrt{\frac{\Lambda}{3}}(r+ \tau))     ).
\end{equation}

For   $ R \gg   a$ and $R \gg   M $ we will have

\begin{equation}
\label{ }
\begin{split}
\tau + r =  &  \int \frac{dR}{\sqrt{\frac{   \Lambda}{3}R^{2}  + \frac{2MR}{R^{2}+a^{2}}    }}   
    \simeq      \int \frac{      (1+\frac{a^{2}}{2 R^{2} }) dR    }{\sqrt{\frac{   \Lambda}{3}R^{2}  + \frac{2M}{R}   }  }       
         \\ &   =      \sqrt{\frac{3}{\Lambda}}  [ ln(R)  + \frac{M}{\Lambda R^{3}} ]  +   \frac{a^{2}}{2}\sqrt{\frac{3}{\Lambda}} [ - \frac{1}{2R^{2}}  +  \frac{3M}{ 5 \Lambda R^{5}}],
\end{split}
\end{equation}

Since   $ R \gg   a$ and $R \gg   M $,  if we neglect $ \frac{1}{R^{5}} $ term, then we will have  $R =  e^{x_{i}} $, where $x_{i}$ are the roots of equation
  \begin{equation}
\label{ }
  -4 \sqrt{3} \Lambda^{\frac{3}{2}} (r + \tau )e^{3x}   + 12 \Lambda x e^{3x}   - 3 \Lambda a^{2} e^{x} +12M =0
\end{equation}

If we only suppose   $ R \gg   a$   (not $ R \gg M  $) we will have

\begin{equation}
\label{ }
\begin{split}
\tau + r =  &  \int \frac{dR}{\sqrt{\frac{   \Lambda}{3}R^{2}  + \frac{2Mr}{R^{2}+a^{2}}    }}      
\simeq      \int \frac{      (1+\frac{a^{2}}{2 R^{2} }) dR    }{\sqrt{\frac{   \Lambda}{3}R^{2}  + \frac{2M}{R}   }  }       
\end{split}.
\end{equation}

Hence if we define 

\begin{align}
\label{ }
& G(R)=  \tau + r =   \sqrt{\frac{4}{3  \Lambda}} arcsinh( \sqrt{\frac{3 \Lambda}{8 M}} \frac{2}{3} R^{\frac{3}{2}} )   +   \\ \nonumber 
&\frac{1}{4 a^{2}}(\frac{3}{2})^{-\frac{2}{3}} \frac{-3 (\frac{  - \Lambda  ( \frac{4}{9} R^{3})}{M})^{\frac{5}{6}} (8M + 3 \Lambda ( \frac{4}{9} R^{3}))   +  2 *3^{\frac{1}{6}} \Lambda  ( \frac{4}{9} R^{3})  \sqrt{8 + \frac{3 \Lambda  (  \frac{4}{9} R^{3})}{M}} Beta[ -\frac{3 \Lambda   \frac{4}{9} R^{3}}{M},\frac{5}{6}   , \frac{1}{2} ] }{  2M ( \frac{4}{9} R^{3})^{\frac{1}{6}}  (\frac{ - \Lambda  ( \frac{4}{9} R^{3})}{M})^{\frac{5}{6}} \sqrt{(8M + 3 \Lambda (  \frac{4}{9} R^{3}))}  },
\end{align}
 
 where $Beta[x,a,b]$ is incomplete beta function. Then we get
 
\begin{equation}
\label{ }
R = G^{-1}(r+\tau ).
\end{equation} 
Therefore, we present point mass CBH with angular momentum with finding the $G$ function.
\\

  The Penrose-Carter diagram of these black holes can be seen in the \cite{kerrdesitter}. 
  \\

\section{Cosmological black holes properties}

From the last sections discussion a question arises as to whether we can see the trace of the cosmological constant (which play the cosmological acceleration role) in the observation. To make an estimate consider the Schwarzschild-de Sitter metric in the stationary coordinate (\ref{s-d}). Since the $R$ value is invariant due to the transformation to the cosmological coordinate, we can compare the two terms $ \frac{\Lambda}{3}R^{2}$ from cosmological constant and $\frac{2M}{R}$ from the black hole. In cosmology the value $ \frac{\Lambda}{3}=H_0^2$ where the $H_0= 67.74 \pm 0.46 km/s / Mpc = 2.195 \pm 0.015 10^{-18} s^{-1}$ is the Hubble constant at the present time. For a black hole with the sun mass $m_\odot$ at scale $ R \gtrsim 10^{18} m \simeq 100 pc $ the cosmological constant term cannot be negligible. In this scale some phenomena of black hole like lensing \cite{mojahed} and Cosmic Microwave Background distortion from primordial black hole are important and we have to use the cosmological constant model to describe them. In this part we investigate what happen for black hole mass, boundary and redshift structure if someone take the cosmological coordinate.
\\

\subsection{Misner-Sharp mass}
The spherically symmetric metric can be written
\be
\label{normal-metric}
ds^2=\gamma_{ab}dx^a dx^b+R^2 d\Omega^2.
\ee
where $a,b: (t,r)$.  In this form the Misner-Sharp mass which is defined for the spherically symmetric space time \cite{Firouzjaee:2010ia}
becomes
\be
M_{ms}=\frac{R}{2}(1-\gamma^{ab}\partial_a R \partial_b R)
\ee 
Since $\gamma^{ab}\partial_a R \partial_b R$ is invariant under the coordinate transformation $(t,r) \rightarrow (t',r')$ the Misner-Sharp mass is also invariant under these transformation.\\
Since the coordinate transformation from the stationary Schwarzschild-de Sitter coordinate to the cosmological coordinate has only the $(t,r)$ part, as a result, spherically symmetric cosmological black holes the Misner-Sharp will be the same Schwarzschild charge de Sitter mass:
\be
M_{ms}=\frac{R}{2}(1-\gamma^{ab}\partial_a R \partial_b R)=m+\dfrac{\Lambda}{6}R^2-\dfrac{Q}{2R}.
\ee 
\\
One can also calculate the matter flux for above cosmological black hole in the spherically symmetric case. It can be shown from equation (\ref{flux}) that the matter flux is zero for these CBHs. This verifies that these CBH are point mass CBHs.
\\

\subsection{Redshift}

If an emitter sends a light ray to an observer with null vector
$k^{\mu}$, the relative light redshift of the emitted ($e$) frequency, ($w$), that is calculated by
observer ($o$) with 4-velocity $u^{\mu}$ is,

\begin{equation}
1+z=\frac{w_e}{w_o}=\frac{(k_\mu u^\mu)_e}{(k_\mu u^\mu)_o},
\end{equation} 
where the light null geodesic $k^{\mu}$ is  affinary parameterized.
It can be shown that the affinnary null geodesics equation 
\be
k^{\nu}k^{\mu}_{;\nu}=0
\ee
will not change from coordinate transformation (because this relation is a covariant tensorial relation).  As a result, the affine parameterized equation will remain the same. Hence, it results that the redshift properties of a spacetime for an observer will not change due to the coordinate transformation.\\
Consequently, the infinite redshift surface for the above CBHs will be the same  infinite redshift surface in the stationary coordinate.\\

\subsection{Black hole boundary}

The event horizon is usually used to define the black hole boundary in  textbooks. Since the event horizon definition is a global quantity and we have to know all information about the spacetime evolution finally, we need to define a quasi-local quantity which can be applied for real dynamical black hole in numerical relativity. To solve this problem people use the isolated on dynamical horizon (which are a type of the apparent horizon) to define the black hole boundary in a dynamical case \cite{Ashtekar:2002ag,Firouzjaee-penn,Booth:2003ji}.\\

\textbf{Black hole boundary definition}: A smooth, three-dimensional, space-like sub-manifold (possibly with boundary) $H  $ of space-time is said to be a trapping horizon  if it can be foliated by a family of closed 2-manifolds such that  on each leaf $S$ the expansion $\theta_{(\ell)}$ of one null normal $\ell^\mu$  vanishes; and the expansion $\theta_{(n)}<0$ of the other null normal $n^\mu$ is negative. This surface separates the trapped surface, $\theta_{(n)}, \theta_{(\ell)}<0$, from untrapped one $\theta_{(n)}<0,~ \theta_{(\ell)}>0$.  Similarly, one can define the cosmological horizon as a three-dimensional surface where the expansion $\theta_{(n)}$ of one null normal $n^\mu$  vanishes and $\theta_{(\ell)}>0$ on both sides of this surface.\\

In the spherically symmetric space time  the black hole boundary is located on the apparent horizon $\theta_{(\ell)}=0$. In terms of normal metric components it becomes $\gamma^{ab}\partial_a R \partial_b R=0$.\\

Since $\gamma^{ab}\partial_a R \partial_b R$ is invariant under the coordinate transformation $(t,r) \rightarrow (t',r')$ the black hole boundary or apparent horizon is  invariant under these transformation.\\

A spacetime with a point mass CBHs have two horizons. The first is cosmological horizon and the second horizon is black hole horizon (event or apparent horizon). For the point mass CBH \cite{Gibbons:1977mu} there are also two black hole and cosmological horizon. Since the coordinate transformation do not change the horizons location, it is sufficient to find  roots of $ \Phi$ in the Schwarzschild-de Sitter spacetime. Note that the infinite redshift surface is the same black hole horizon surface in the stationary coordinate. Hence, the infinite redshift surface will be the same black hole apparent horizon for these CBHs. It can easily be shown that the expansion of outgoing null geodesics is proportional to $\Phi $, which $\Phi =0 $ determines the position of  horizons. We can  depict the roots of $ \Phi = 1- \frac{\Lambda}{3}R^{2} - \frac{2m}{R}  $ for $ \Lambda  = 10^{-52}m^{-2} $ as a function of $x=\frac{m}{m_{\odot}} $ where $m_{\odot}$ is the sun mass. There are three real roots for  $\Phi=0 $ if $ \Lambda m^{2} \leqslant \frac{1}{9} $ \cite{Guven}, where one of them is negative, so we consider $R_{C} = \frac{2}{\sqrt{\Lambda}} \cos(\frac{\cos^{-1}(-3 \sqrt{\Lambda}m)}{3})$ and $R_{H} =   \frac{2}{\sqrt{\Lambda}} \cos(\frac{\cos^{-1}(-3 \sqrt{\Lambda}m)}{3} + \frac{4 \pi}{3})  $ which are  cosmological and black hole  event horizons respectively.
  
\begin{figure}[h] 		\centering \includegraphics[width=0.4 \columnwidth]{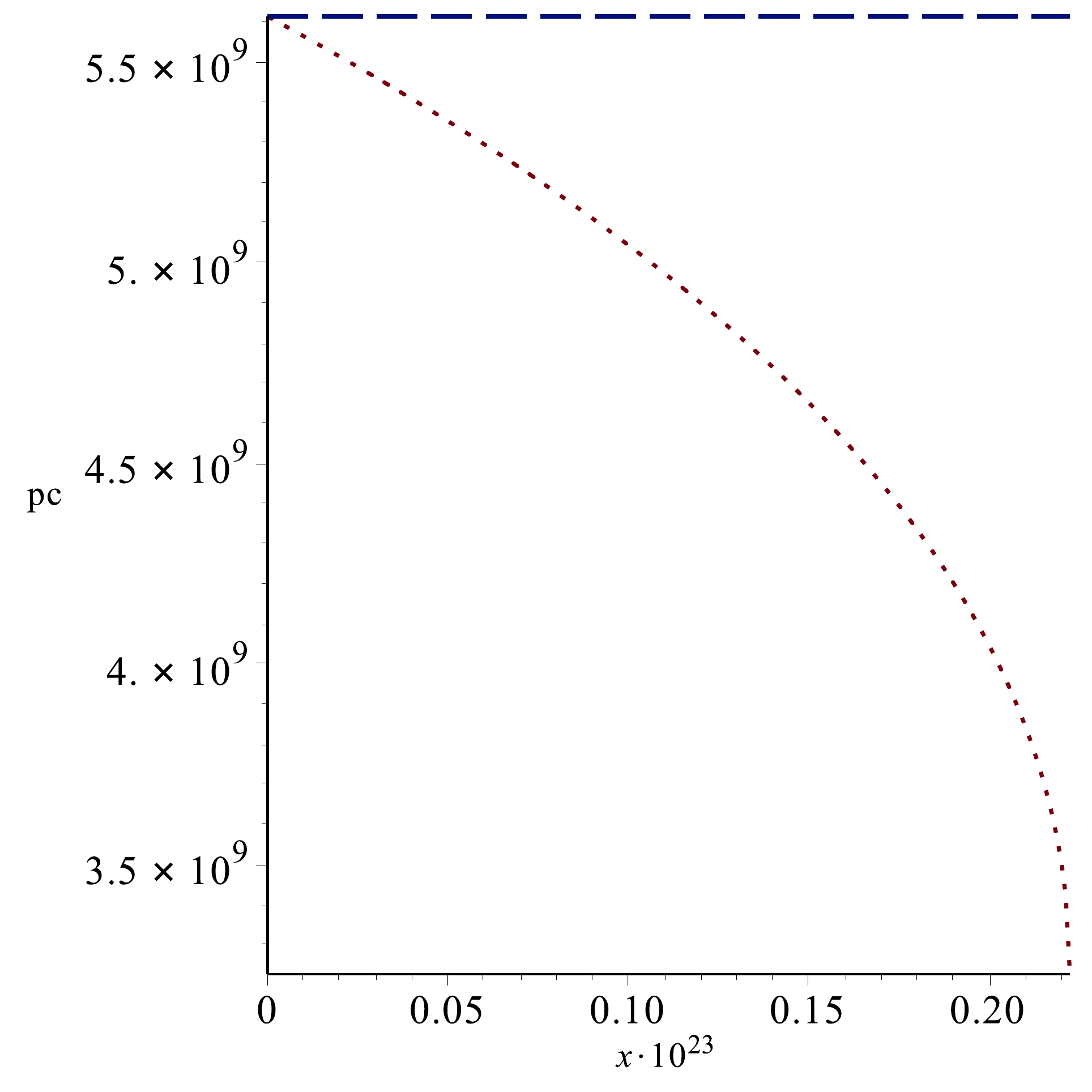}  \caption{de Sitter event horizon as a function of mass. Dash line represents de Sitter event horizon for $m=0$.} 	\label{fig:10}	
\end{figure}
\begin{figure}[t]
	\centering \includegraphics[width=0.4 \columnwidth]{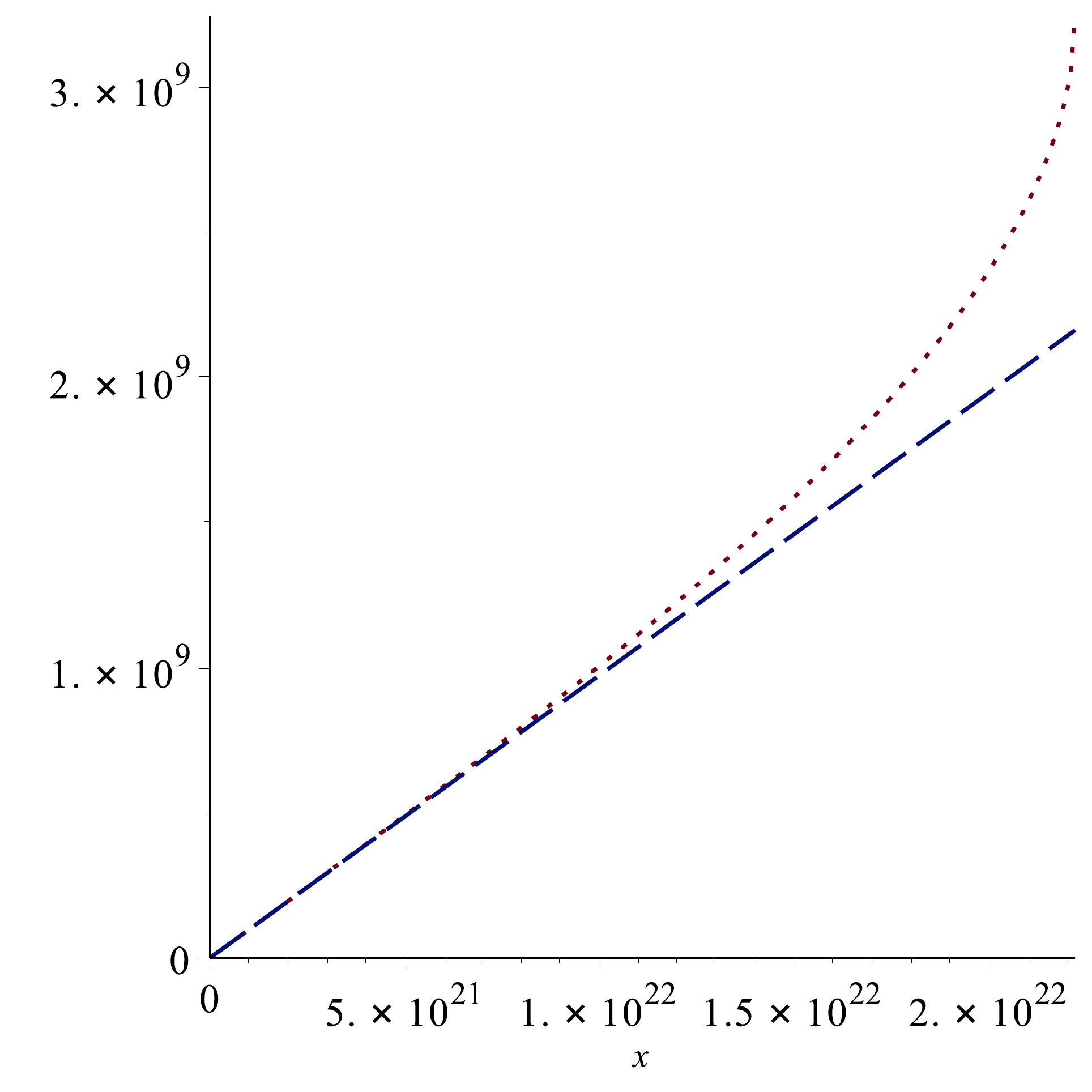} 
	\caption{Black hole event horizon as a function of mass. Dash line represents black hole event horizon for $\Lambda=0$.}\label{fig:11} 		
\end{figure}

Two cosmological and black hole horizons are depicted in Fig.(\ref{fig:10}) and Fig.(\ref{fig:11}).

 It can be seen that using standard value of dark energy, the astrophysical black hole with mass $m < 10^{10} m_\odot$ can not change the cosmological horizon place. On the other side, the standard value of cosmological constant does not change the black hole horizon place significantly.\\

\subsection{Circular orbits for CBHs}
    
    In cosmology and astrophysics, many events and physics such as microlensing, dark matter rotation curve, missing satellite, stars velocity dispersion and etc. Emerge from the studying of object orbits  where most of them are simplified by circular orbits \cite{desitter-geo}. In this part, we study the possibility of the circular orbits around the CBHs (without charge and angular momentum). \\

    Circular orbits are characterized by a constant radius. In a stationary coordinate like Schwarzschild de Sitter coordinate there is well defined coordinate $R$ which is the areal coordinate, but in the cosmological coordinate we have two comoving and areal radius. Since the angular distance is proportional to the areal radius we  choose the areal radius as a circular orbits radius. As a consequence of time and radius coordinate transformation  from stationary Schwarzschild-de Sitter to the CBH cosmological coordinate, the areal coordinate will not changes. As a result, the circular coordinate in both frames are the same, but are labeled by different coordinates. The detail of the orbits effective potential can be seen in the Appendix A.\\

\section{Conclusion}

Real black holes in the universe, called the cosmological black holes, are located in the expanding accelerating background. These black holes are generically dynamical and they are sourced by baryons and dark matter. It's been shown that the background expansion leads to voids. These voids are formed between the black hole and the expanding background and prevent the black hole's matter flux from increasing. After this phase of the black hole evolution, the black holes can be approximated as a point mass. In this case, the black hole mass is much greater than the matter flux around it. In this paper, we argue that most of the dynamical cosmological black holes can be modeled by point mass cosmological black holes finally. The point mass cosmological background is modeled by the de Sitter space-time. We find the  Schwarzschild-de Sitter metric in the cosmological coordinate and infer that this metric is the best candidate for the point mass CBH. This metric locally describes a point mass black hole and at large distance reduces to the cosmological de Sitter space-time. We also find the point mass black holes solution with charge and angular momentum. We show that the mass, horizons and redshift structure of these black holes will not change due to the coordinate transformation from stationary coordinate to the cosmological coordinate. From studying the effective potential for different geodesics cases it has been shown that the stable circular orbits can be exist similar to the stationary case.
\\

Future work \cite{Fagihi} will concentrate on studying  the astrophysical and cosmological observables of these black holes and compare them with stationary ones.
\\

{\bf Acknowledgments:}
\\

We would like to thank Alireza Allahyari for useful
comments.
\\
\appendix
\section{The CBH's orbits effective potential }
Here we want to study the  form of effective potential  in Schwarzschild-de Sitter spacetime to study the circular orbits. We can use Euler Lagrange Equations with metric (\ref{a1})  which is  spherically symmetric and so we can choose $\theta = \frac{\pi}{2}$, hence  $ d \Omega ^{2} = d \varphi ^{2} $ \cite{circular}. The Euler Lagrange equations will result:       
\begin{equation}
\label{z1}
\begin{split}
& \epsilon =  -\Phi \dot{t}^{2} + \Phi^{-1} \dot{R}^{2} + R^{2} \dot{ \varphi } ^{2}
\\ & \dot{t} = \frac{E}{\Phi}
\\&  \dot{ \varphi } = \frac{L}{R^{2}} ,
\end{split}
\end{equation}
where dot  is partial differential relative to affine parameter  and E and L are constants. $\epsilon = -1,0 ,1$ represents timelike, null and spacelike geodesics respectively.
From the equation (\ref{z1}) we can introduce effective potential V(R), where 
\begin{equation}
\label{z5}
V(R) = E^{2} -  \dot{R}^{2} = \Phi(R)(\frac{L^{2}}{R^{2}} - \epsilon) 
\end{equation} 
We can depict V(R) as a function of $ \frac{M}{R}$ for different values of $L$ and  $ \alpha  $ where $ \alpha = 9\Lambda M^{2} $ and $ 0 \leqslant \alpha  \leqslant 1  $, that $\alpha=1 $ is the extreme case \cite{podolski}.

\begin{figure}[h]
	\centering
	\begin{minipage}[b]{0.4\textwidth}
		\centering
		\includegraphics [width=1.3\linewidth] {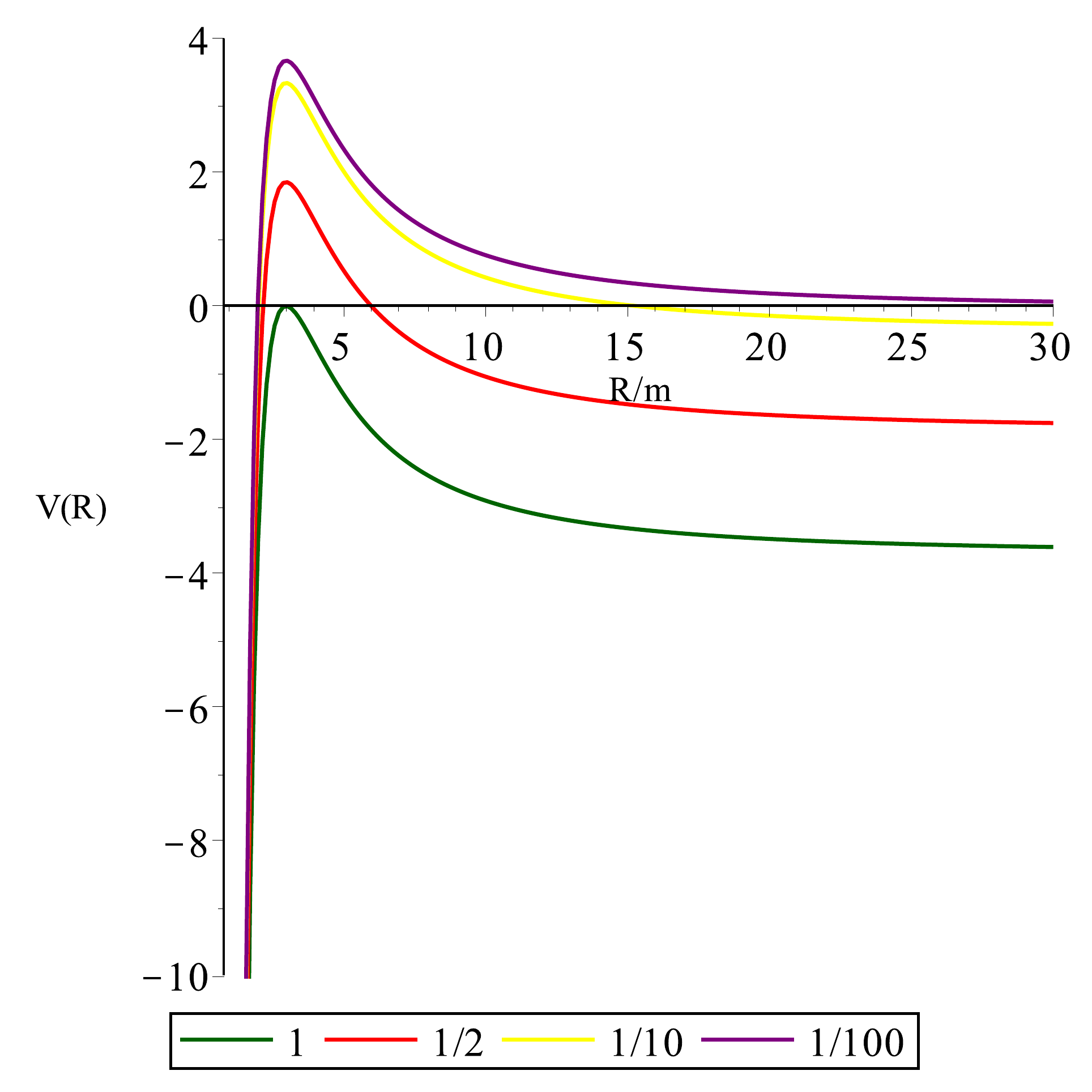}
		\caption{ Effective potential of null geodesics as a function of $\frac{R}{M} $ for different values of $ \alpha$. Here we have supposed  $L=10M $}
		\label{zz1}
	\end{minipage}
	\hfill
	\begin{minipage}[b]{0.4\textwidth}
		\centering
		\includegraphics [width=1.3\linewidth ]{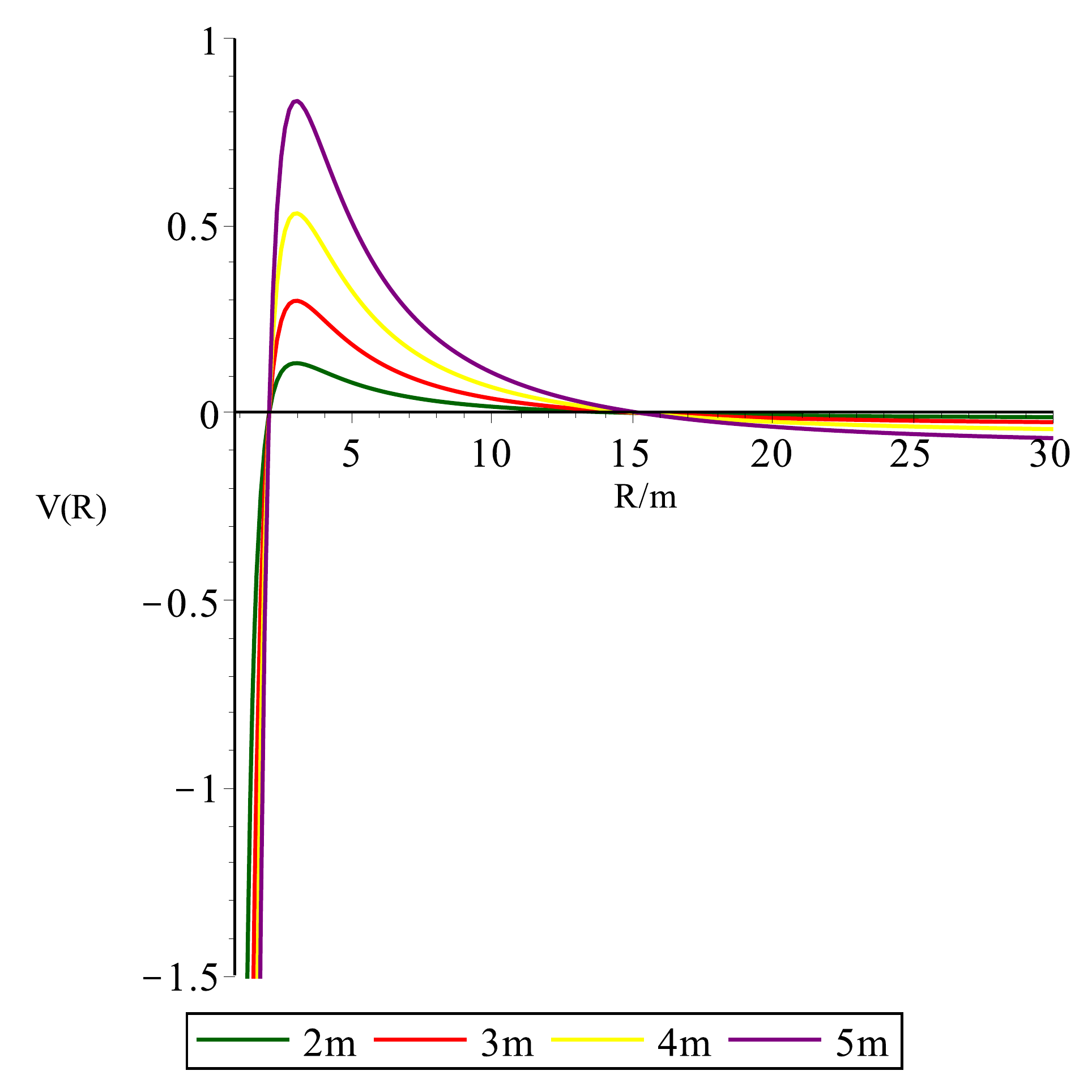}
		\caption{ Effective potential of null geodesics  as a function of $\frac{R}{M} $ for different values of L. Here we have supposed  $\alpha=\frac{1}{10} $}
		\label{zz2}
	\end{minipage}
\end{figure}

If we suppose a massless particle at infinity has the initial velocity $v^{R}_{0} = \sqrt{E^{2} + V(\infty)}$, as the massless particle moves toward central object, it decelerates to $ \sqrt{E^{2}-V(R)} $ and if $  E^{2} > V_{max}$, the zero mass particle will accelerate toward singularity otherwise $\dot{r}$ will change sign and the particle will be reflected to infinity.   According to figure (\ref{zz1}) and (\ref{zz2}) as we decrease $\Lambda$ or increase L, it becomes harder for these particles to fall in to singularity.  We can also study the circular orbits for mass less particles. We know that  if $\frac{\partial V}{\partial R} =0 $  and $ \dot{R}=0$, then we will have circular orbits and the stability criterion is $\frac{\partial ^{2} V}{\partial R^{2}} > 0 $. Hence we can write

\begin{equation}
\label{cv1}
\begin{split}
& \frac{\partial V}{\partial R} =0 \rightarrow R= 3M
\\& \dot{R} = 0   \rightarrow E^{2} = V(R).
\end{split}
\end{equation}
which has the same radius ( $R= 3M$) as the $\Lambda = 0 $ case \cite{bh-book}.  

According to figure (\ref{zz1}), $(\frac{\partial ^{2} V}{\partial R^{2}})_{| R=3M}  < 0 $, that means there is no stable circular orbit for mass less particles except when $ E=L =0$ and $R= 3M$. By equation (\ref{cv1}), if we define  critical  impact parameter as $b= \frac{L}{E}$, then  we can write:
\begin{equation}
\label{ }
b_{critic} =\frac{L}{E} = (\sqrt{\frac{R^{2}}{\Phi(R)}})_{| R=3M} = \frac{3 \sqrt{3} M}{\sqrt{1-\alpha }}.
\end{equation}
If the impact parameter of these mass less  particles is greater than critical value ($b_{critic}$), then these particles will be captured.  This definition leads to a relation for capture cross section for massless particles from infinity:
\begin{equation}
\label{ }
\sigma_{null} = \pi b^{2} = \frac{27 \pi M^{2}}{1- \alpha}.
\end{equation}

\begin{figure}[h]
	\centering
	\begin{minipage}[b]{0.4\textwidth}
		\centering
		\includegraphics [width=1.3 \linewidth] {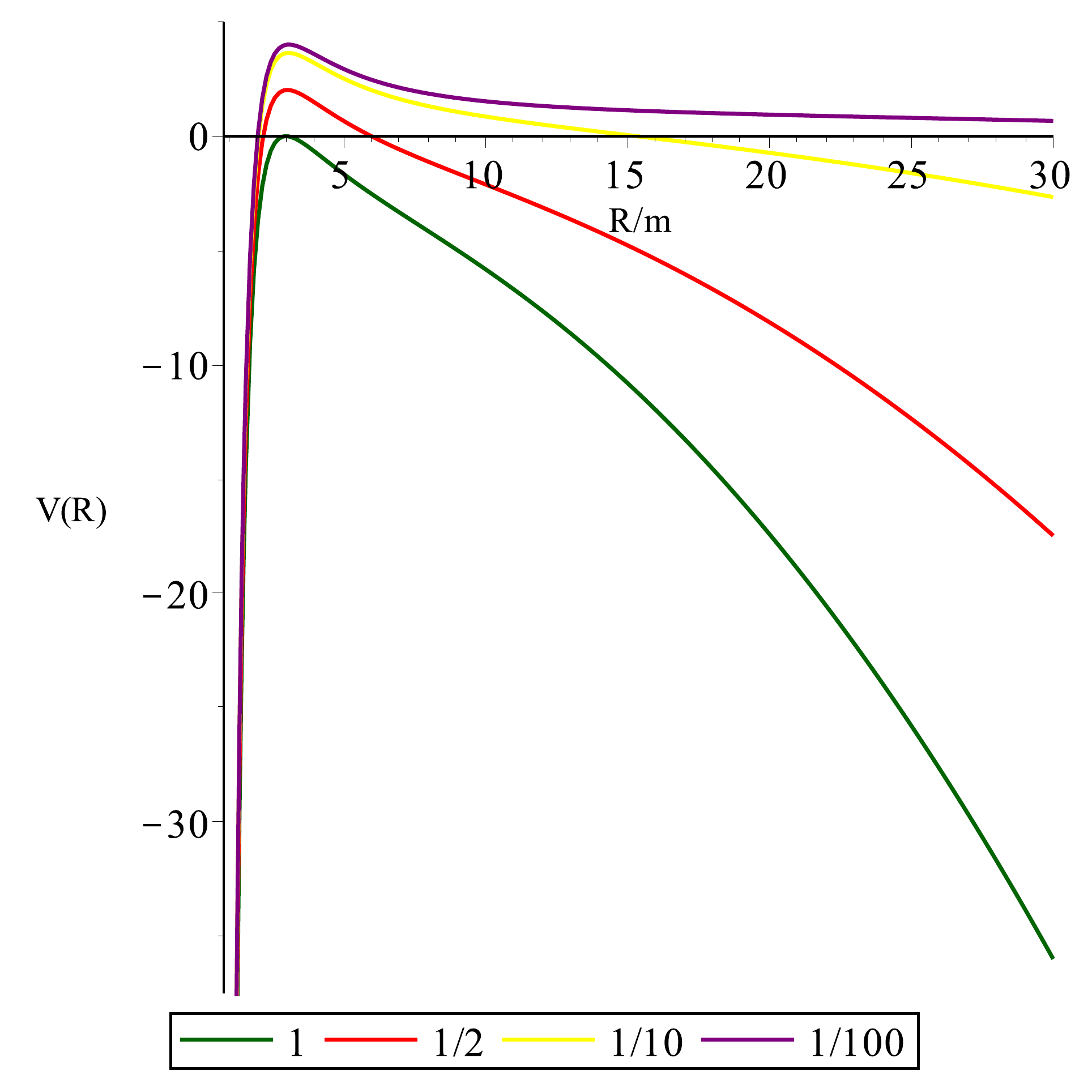}
		\caption{ Effective potential of timelike geodesics as a function of $\frac{R}{M} $ for different values of $ \alpha$. Here we have supposed  $L=10M $.}
		\label{zz3}
	\end{minipage}
	\hfill
	\begin{minipage}[b]{0.4\textwidth}
		\centering
		\includegraphics [width=1.3\linewidth ]{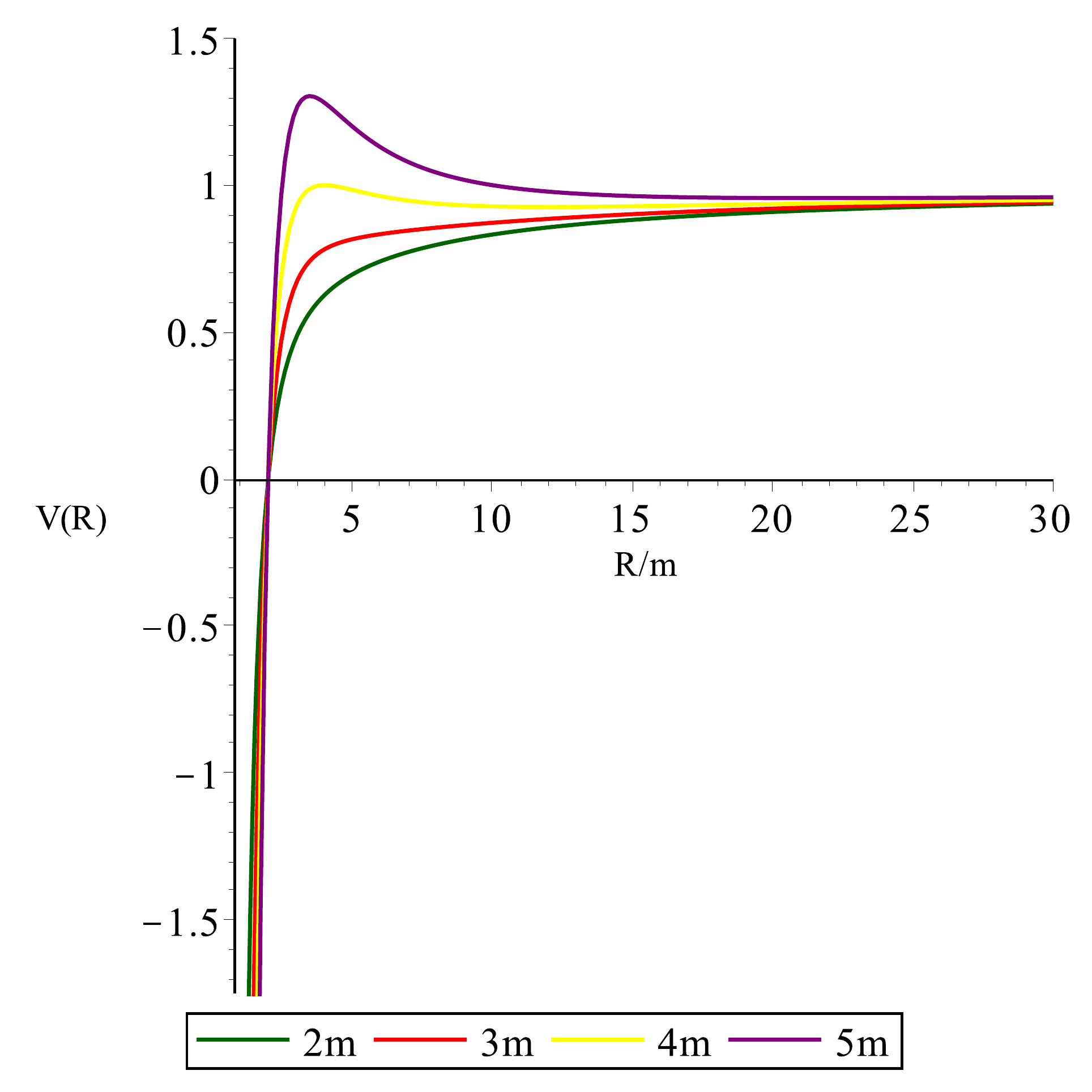}
		\caption{ Effective potential of timelike geodesics  as a function of $\frac{R}{M} $ for different values of L. Here we have supposed  $\alpha=\frac{1}{10} $.}
		\label{zz4}
	\end{minipage}
\end{figure}

\begin{figure}[h]
	\centering
	\begin{minipage}[b]{0.4\textwidth}
		\centering
		\includegraphics [width=1.3\linewidth] {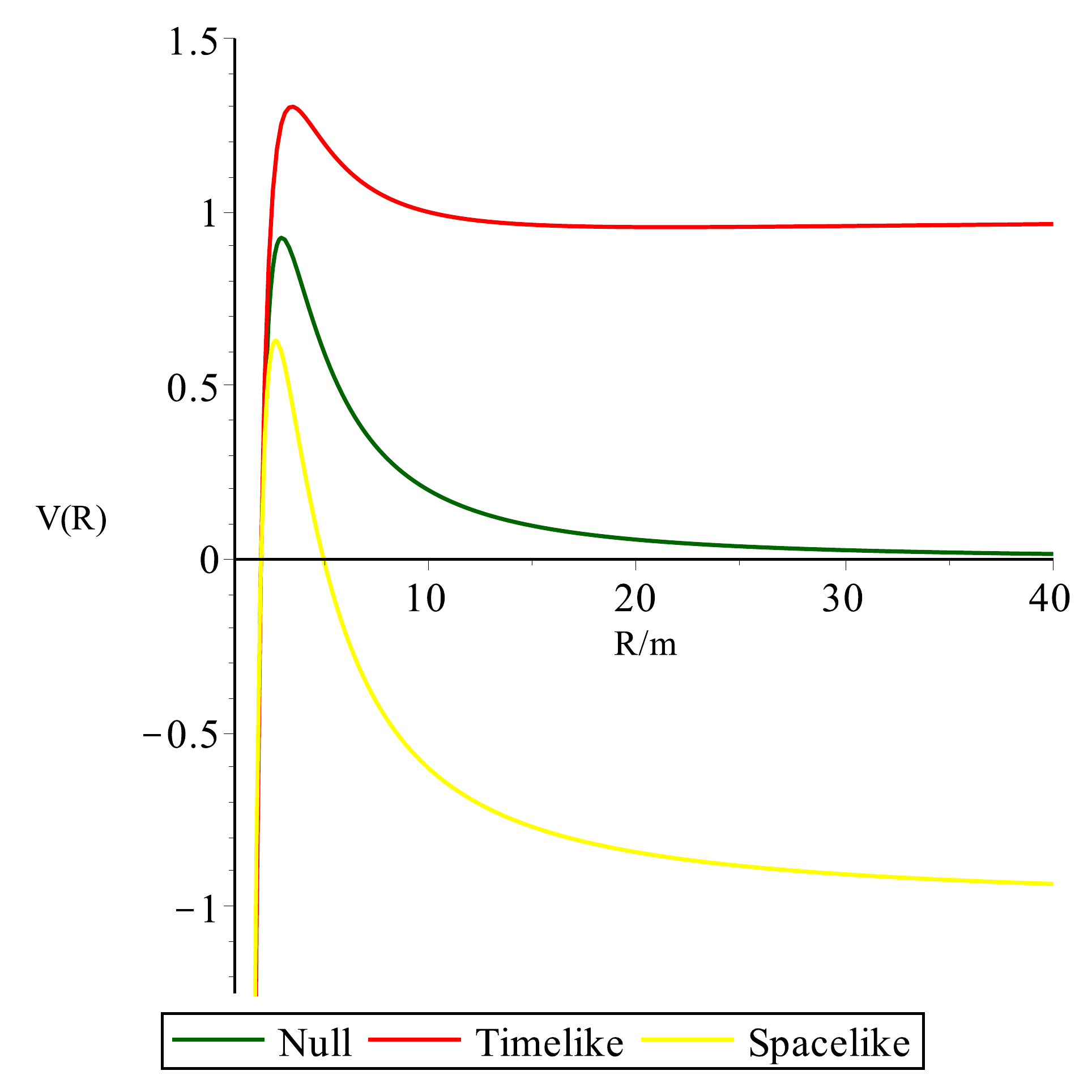}
		\caption{ Effective potential of spacelike,null and timelike  geodesics as a function of $\frac{R}{M} $ for a black hole with mass $10^{6}M_{\odot}$ and $\Lambda = 10^{-52}s^{-1}$. Here we have supposed  $L=5M $.}
		\label{cx1}
	\end{minipage}
	\hfill
	\begin{minipage}[b]{0.4\textwidth}
		\centering
		\includegraphics [width=1.3\linewidth ] {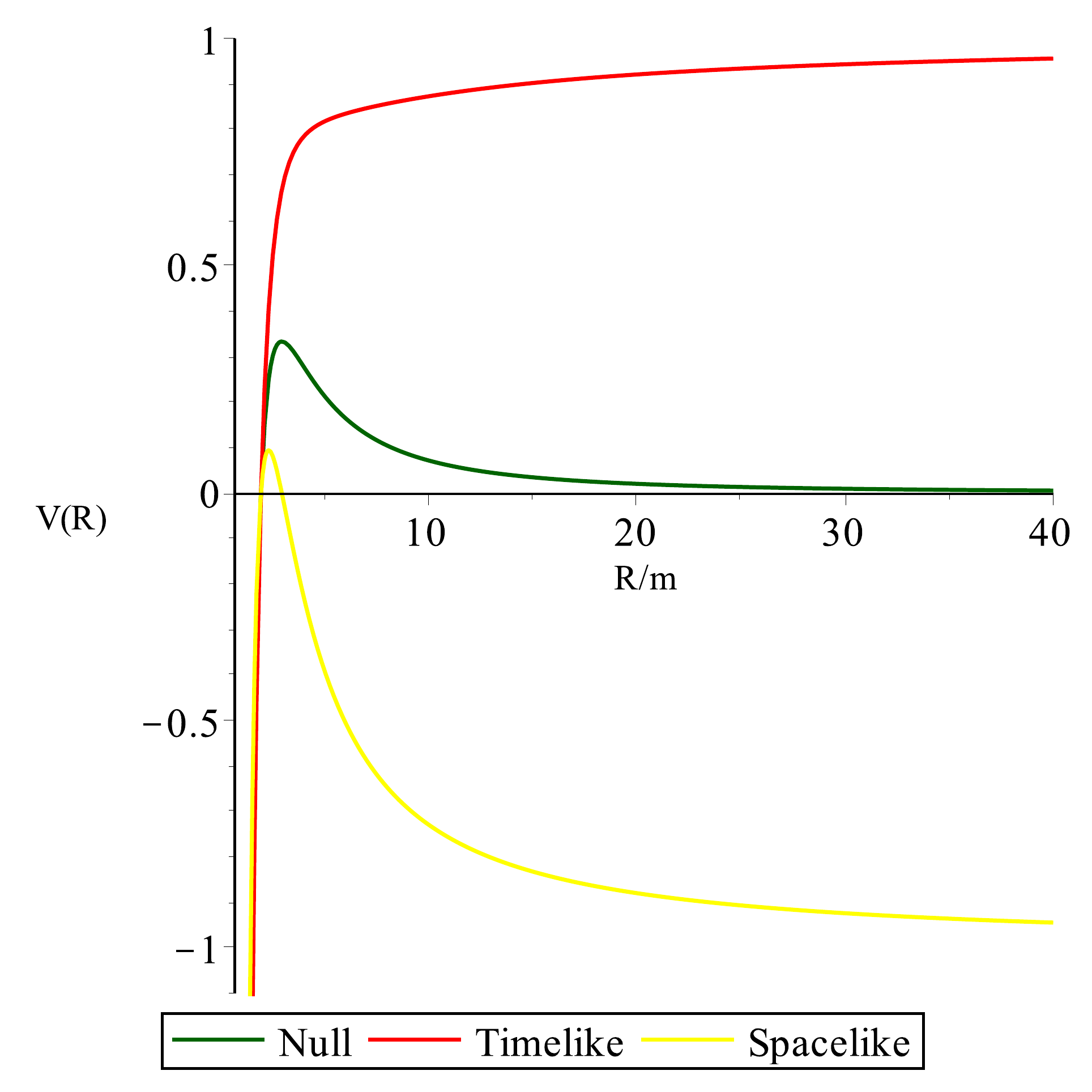}
		\caption{ Effective potential of spacelike,null and timelike  geodesics as a function of $\frac{R}{M} $ for  a black hole with mass $10^{6}M_{\odot}$ and $\Lambda = 10^{-52}s^{-1}$. Here we have supposed  $L=3M $.}
		\label{cx2}\end{minipage}
\end{figure}

We can do the same calculations for the timelike circular orbits. In this case we have

\begin{equation}
\label{cv10}
\frac{\partial V}{\partial R} = 0 \rightarrow L^{2} = \frac{MR^{2} - \frac{\Lambda}{3}R^{5} }{R - 3M}.
\end{equation} 
By equation (\ref{cv10}) and $ \dot{R} = 0$  one can easily show: 
\begin{equation}
\label{cv13}
E^{2}=  V (R)=\frac{(1-\frac{\alpha}{27M^{2}}R^{2} -\frac{2M}{R} )^{2}}{R(R -  3M)}.
\end{equation}
According to equation (\ref{cv13}) it's clear that circular orbits exist for $ R > 3M $, but figure (\ref{zz3}) tells us that effective potential is convex for the given values of cosmological constant, so there is no circular orbit for time like case at these values.  The value of $\alpha$ for  a black hole with mass $10^{6}$ sun mass and $\Lambda = 10^{-52} s^{-1}$ is in order of $10^{-32}$, in this case we can have stable circular orbit and  bound orbit ( figure (\ref{cx1}) ).    
Figure (\ref{zz3}) also  represents that    for a same value of central mass if we increase the cosmological constant, $\dot{R}$ increases. For a same value of $\Lambda $, figure (\ref{zz4}) shows that any variation of L in the timelike case is similar to null one.  

\begin{figure}[h]
	\centering
	\begin{minipage}[b]{0.4\textwidth}
		\centering
		\includegraphics [width=1.3\linewidth] {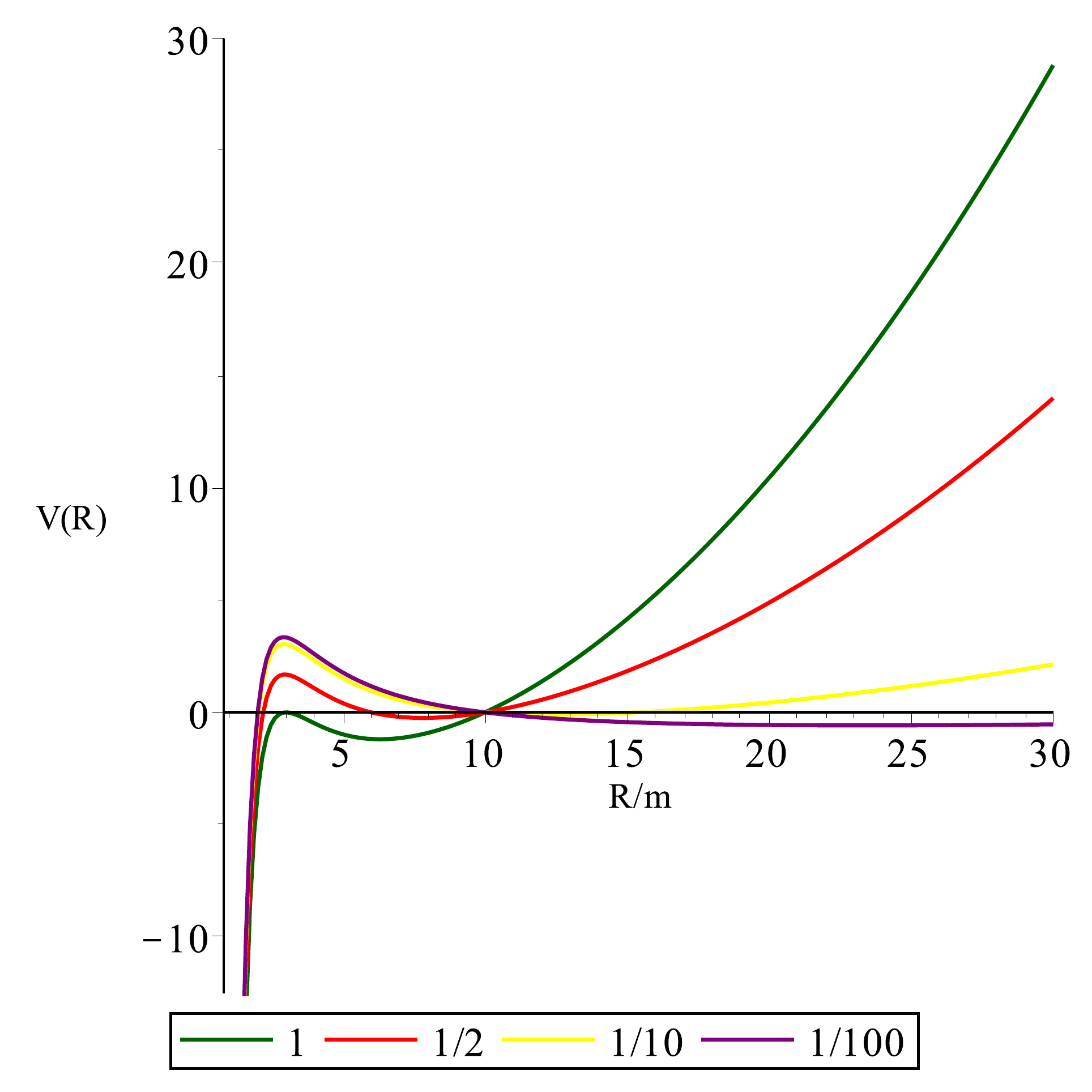}
		\caption{ Effective potential of spacelike geodesics as a function of $\frac{R}{M} $ for different values of $ \alpha$. Here we have supposed  $L=10M $.}
		\label{zz5}
	\end{minipage}
	\hfill
	\begin{minipage}[b]{0.4\textwidth}
		\centering
		\includegraphics [width=1.3\linewidth ]{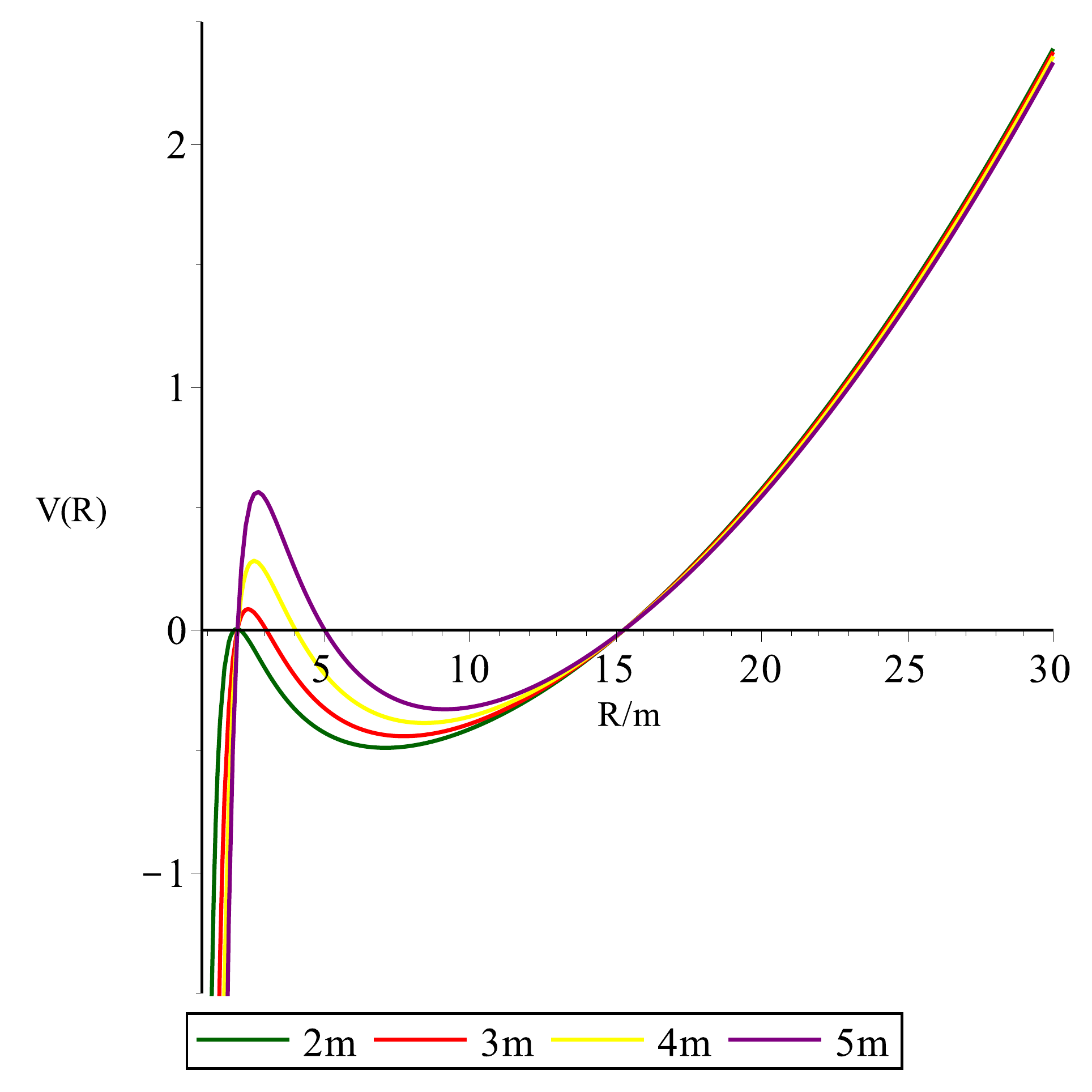}
		\caption{ Effective potential of spacelike geodesics  as a function of $\frac{R}{M} $ for different values of L. Here we have supposed  $\alpha=\frac{1}{10} $.}
		\label{zz6}
	\end{minipage}
\end{figure}

Figure  (\ref{zz5}) shows that the spacelike case  inside the de Sitter horizon resembles  the null and timelike cases, but for outside de sitter horizon as we increase $\Lambda $ (for same vale of central mass), $\dot{R}$ decreases.


Figure (\ref{cx1})  and   (\ref{cx2}) represent effective potential for  a  specific black hole with mass  $10^{6}M_{\odot}$ and $L=5M $ and $3M$. We have also used present value of $\Lambda = 10^{-52}s^{-1}$.   We can see that only for some values of L we can have timelike circular orbits and bound orbits. For the extreme case you can see \cite{podolski}.

\end{document}